\documentclass[multphys,vecphys]{svmult}

\usepackage{makeidx}     
\usepackage{graphicx}    
\usepackage{multicol}    

\makeindex             

\begin{document}

\title*{NUclei of GAlaxies (NUGA): the IRAM Survey of Low Luminosity AGN}
\titlerunning{NUclei of GAlaxies (NUGA)}

\author{S. Garc\'{\i}a-Burillo\inst{1}, F. Combes\inst{2}, A. Eckart\inst{3}, L.~J. Tacconi\inst{4},
L.~K. Hunt\inst{5}, S. Leon\inst{6}, A.~J. Baker\inst{4}, P. Englmaier\inst{7}, F. Boone\inst{8},
E. Schinnerer\inst{9} \and R. Neri\inst{10}}


\authorrunning{Garc\'{\i}a-Burillo et al}


\institute{Observatorio Astron\'omico Nacional (OAN), Madrid, Spain \texttt{s.gburillo@oan.es} \and
Observatoire de Paris, LERMA, Paris, France \texttt{francoise.combes@obspm.fr} \and
Universit\"at zu K\"oln, K\"oln, Germany  \texttt{eckart@ph1.uni-koeln.de} \and
MPE, Garching, Germany \texttt{linda@mpe.mpg.de, ajb@mpe.mpg.de} \and
Instituto di Radioastronomia/CNR,Firenze, Italy \texttt{hunt@arcetri.astro.it} \and
Instituto de Astrof\'{\i}sica de Andaluc\'{\i}a, Granada, Spain   \texttt{stephane@iaa.es} \and
Universit\"at Basel, Binningen, Switzerland  \texttt{ppe@astro.unibas.ch } \and
Bochum University, Bochum, Germany   \texttt{fboone@astro.rurh-uni-bochum.de } \and
NRAO, Socorro, USA  \texttt{eschinne@nrao.edu } \and
IRAM, Grenoble, France  \texttt{neri@iram.fr}}
%
%
\maketitle

The NUGA project (Garc\'{\i}a-Burillo et al \cite{cont1},\cite{paper1}) is the
first high-resolution ($\sim$0.5$^{\prime\prime}$--1$^{\prime\prime}$) CO survey of low luminosity
AGN including the full sequence of activity types (Seyfert 1--2 and LINERs). Other interferometric
surveys had not been focused on the study of AGN thus far (e.g., Helfer et al \cite{hel03}). NUGA
aims to study systematically the different mechanisms for gas fueling of the
AGN. There is evidence that the critical spatial scales for AGN feeding are $<$100~pc. Any causal
link between a particular type of instability and the onset of activity must therefore be sought
in {\it secondary} modes embedded in the kpc-scale perturbations
(bars or spirals). High-resolution observations are thus paramount to achieve a sharp view of the
distribution and kinematics of molecular gas down to the critical scales in AGN. 

NUGA has a multi-wavelength approach: the availability of HST and ground-based optical/NIR images
for
most of the targets allows us to probe the stellar potentials and to study the star formation
patterns in NUGA galaxies. Ongoing VLA (HI) and VLBI (radiocontinuum) NUGA-related projects probe
the large scale disks and the compact nuclear sources. Altogether, this multi-wavelength information
allows us to test the efficiency of AGN feeding mechanisms thanks to a systematic evaluation of the
stellar gravitational torques exerted on the gas and, also, with the help of numerical simulations
of the gas dynamics for the galaxies in our sample.

\begin{figure}
\flushleft
\includegraphics[height=5.85cm]{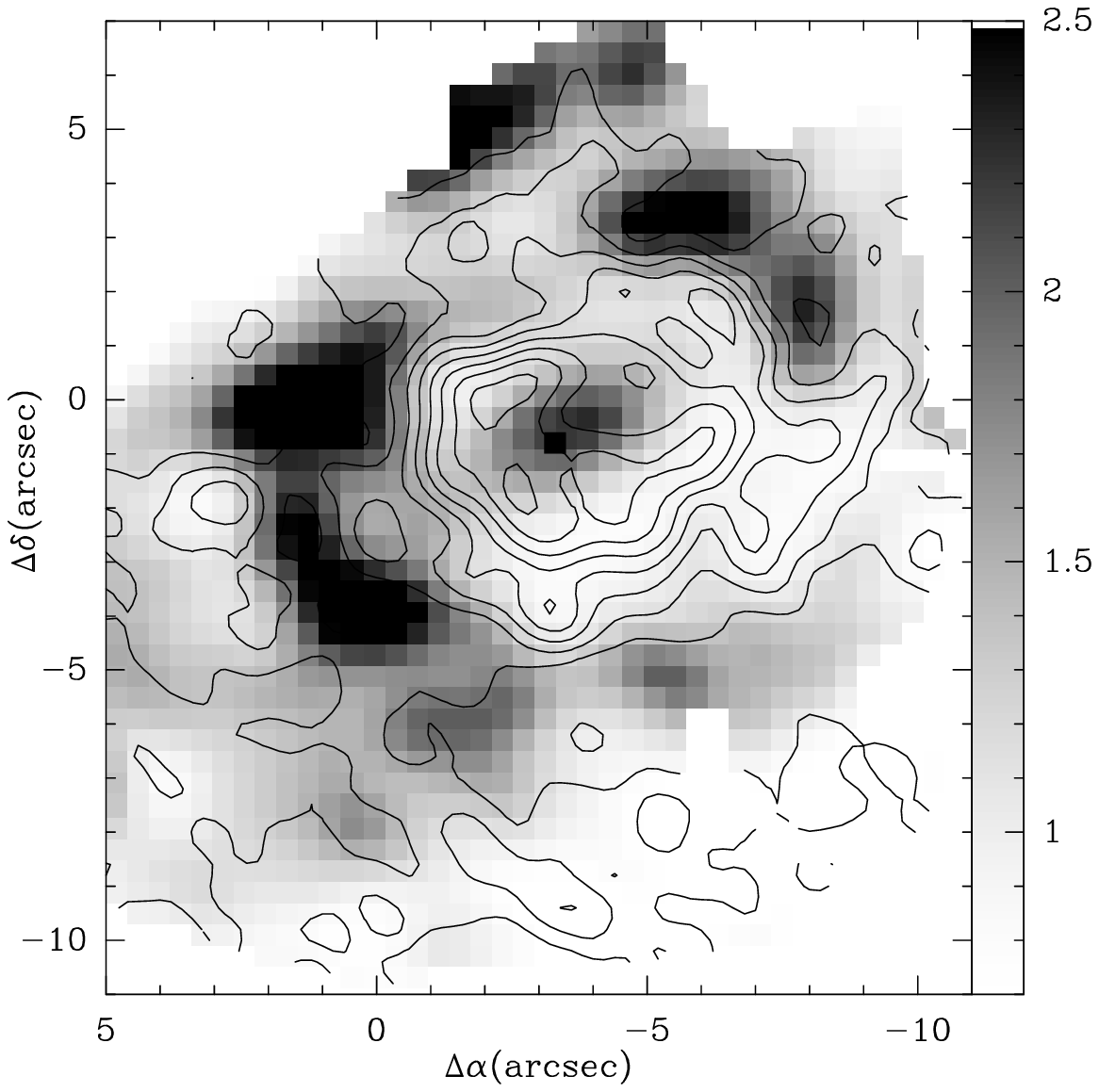}
\flushright
\vspace{-6.cm}
\includegraphics[height=5.45cm, angle=0]{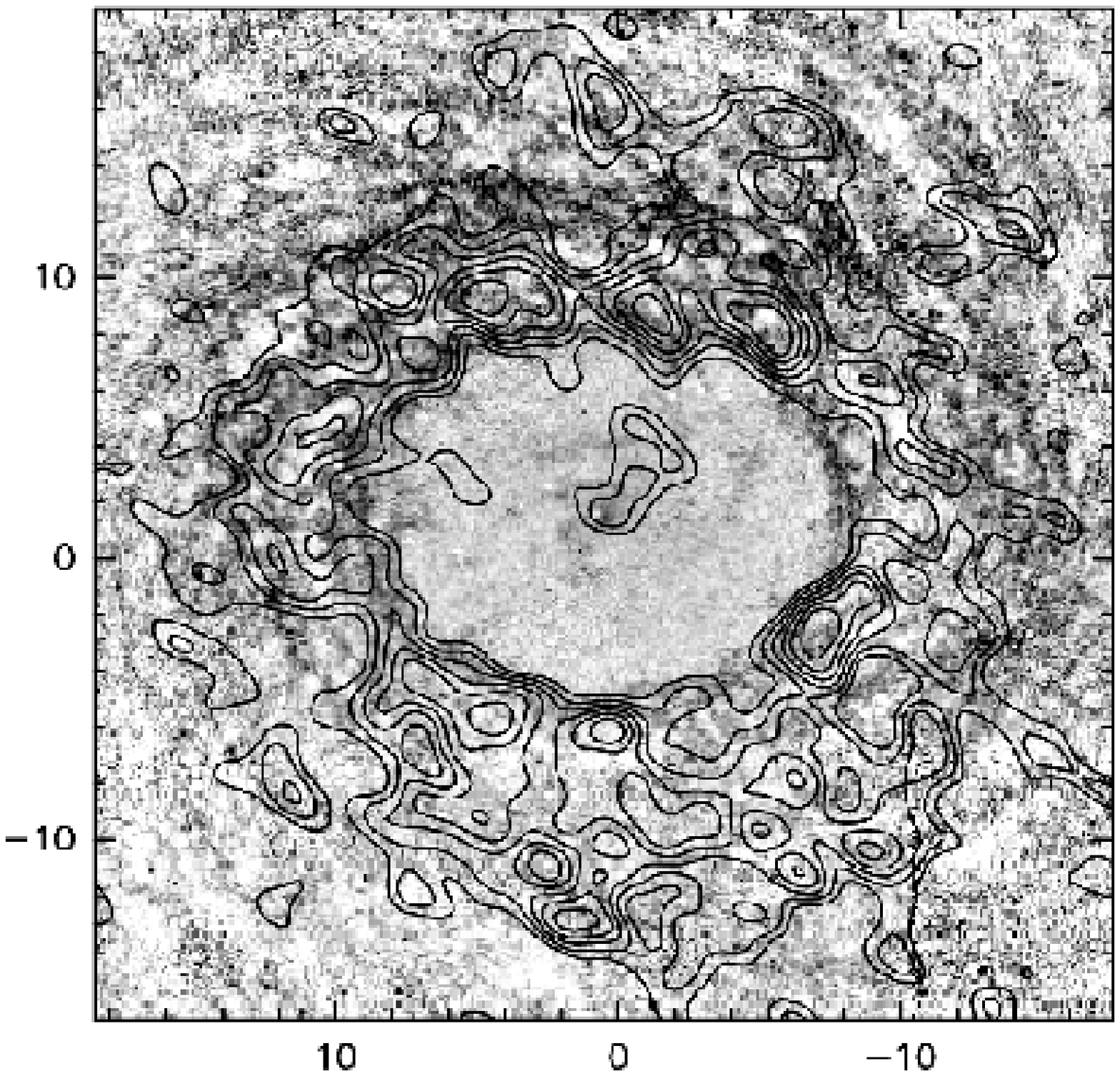}
%
%
\caption{{\bf Left:a)} Map of the Toomre Q parameter (grey scale) overlaid on the
$^{12}$CO(2--1) integrated intensity contours of NGC\,4826 from Garc\'{\i}a-Burillo et al
\cite{paper1}.
Gas self-gravity seems important in the $m$=1 modes identified in NGC\,4826. {\bf Right:b)}
$^{12}$CO(1--0) integrated intensity  contours in NGC\,7217 from Combes et
al \cite{paper2}, overlaid on HST V--I color map. Molecular gas forms a contrasted ring. A low-mass
non self-gravitating gas unit is situated on the AGN, at the center of the
map.}  
\label{Fig:1}       
\end{figure}
%
\begin{figure}
\flushleft
\includegraphics[height=10cm]{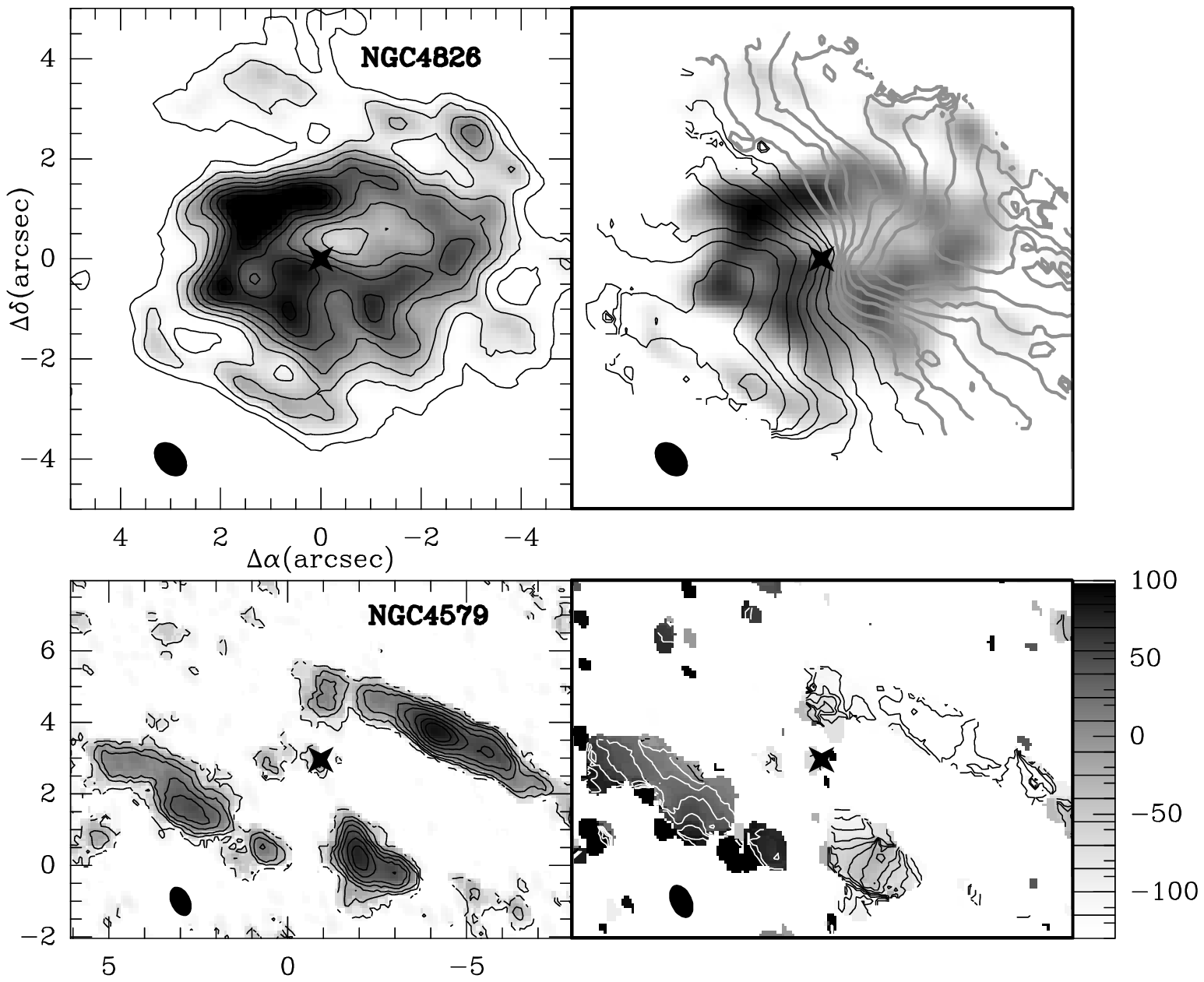}

%
%
\caption{{\bf Top:} High-resolution ($\sim$0.5$^{\prime\prime}$--1$^{\prime\prime}$) new
data gives a sharp view of gas kinematics in the lopsided
circumnuclear disk of NGC\,4826 . The $^{12}$CO(2--1) integrated intensity map is shown in {\it
left} panel (2.5-8.5~Jy~km~s$^{-1}$) and the overlay with velocity contours in {\it right} panel
(-120 to 135 by 15~km~s$^{-1}$). {\bf Bottom:} The $m$=1 mode
identified in  the $^{12}$CO(2--1) map ({\it left}) of NGC\,4579 (scale=0.1 to 9~Jy~km~s$^{-1}$) is
revealed by a strongly asymmetrical velocity gradient detected only on the eastern arm ({\it
right}). Figures taken from  Garc\'{\i}a-Burillo et al (2004), in prep.} 

\label{Fig:2}       
\end{figure}

\section{First Results of NUGA} 
\label{First}
The first CO NUGA maps made with the Plateau de Bure Interferometer reveal a rich variety of
morphologies in the circumnuclear disks of AGN hosts. Various gravitational instabilities have been
identified at different spatial scales,
including $m$=2, $m$=1 and $m$=0 features. Some galaxies host several coexisting perturbations,
while others host mainly one type of instability.

The distributions and kinematics of molecular gas in several galaxies reveal the presence of $m$=1 
perturbations, which appear as one-arm spirals and lopsided disks developing from several tens to
several hundreds of pc. NUGA observations address for the first time the role of $m$=1
instabilities in AGN fueling. Paper I (Garc\'{\i}a-Burillo et al. \cite{paper1})
focused on the counter-rotating LINER NGC\,4826: a prototype for $m$=1 modes. The
BCD-maps of NGC\,4826 show that the molecular gas is distributed in a 40\,pc-radius lopsided
disk and two $m$=1 spirals (Fig.~\ref{Fig:1}). Streaming motions linked with the $m$=1 perturbations
are compatible with the inner modes being {\it fast} trailing waves. This implies that the lopsided
instability may slow down or even temporarily halt gas infall. An independent confirmation that AGN
feeding is inhibited at present in NGC\,4826 has been obtained from an evaluation of the stellar
gravitational torques exerted on the gas.

We have compared the NUGA observations of the ringed
LINER NGC\,7217 with numerical simulations in Paper II (Combes et al. \cite{paper2}). This galaxy is
in direct contrast with what we observe in NGC\,4826: in NGC\,7217
the molecular gas is distributed in an astonishingly axisymmetric ring (Fig.~\ref{Fig:1}). N-body
simulations, including gas and star-formation, monitor the formation and evolution of the ring;
results from these simulations led us to conclude that gas inside the ring, instead of helping AGN
feeding, is experiencing an outward flow. Only a GMC-like non self-gravitating unit is linked with
the AGN at the center of the map.

\section{New Subarcsec Resolution Maps}
\label{New}

The first $\sim$0.5$^{\prime\prime}$--1$^{\prime\prime}$ resolution maps, incorporating
AB-configuration data, have been recently obtained for five galaxies: NGC\,6951, NGC\,4826,
NGC\,4579, NGC\,2782, and NGC\,1961. We briefly report here on new results obtained for two
galaxies.  The new CO(2--1) data for NGC\,4826 (Fig.~\ref{Fig:2}) have provided the first CO-line
probe of the kinematics and distribution of a molecular gas disk near an AGN at a spatial
resolution of $\leq 10\,{\rm pc}$. In the new maps, the inner {\it fast} lopsided instability is
fully resolved; a secondary $m$=2 outer spiral leaves its imprint of {\it slow} trailing
perturbation in the highly perturbed gas kinematics. The combination of $m$=1+$m$=2 modes seems to
cancel out any gas inflow to the AGN at present. New CO(2--1) A-configuration data of NGC\,4579
have fully resolved the asymmetric kinematics of molecular gas (Fig.~\ref{Fig:2}). The new
observations have revealed a transverse velocity gradient on the eastern arm which has surprisingly
no counterpart on the western arm; this is further evidence for the existence of a lopsided
instability. 

\section{Conclusions}
\label{concl.}

The large variety of circumnuclear disk morphologies found in NUGA galaxies is a challenging result
that urges the refinement of current dynamical models. The lack of a clear correlation between
activity type and nuclear morphology of the AGN host suggests that the AGN duty cycle in low-L AGN
is typically shorter than the time scale needed to build up the described gravitational
instabilities at scales of a few ten-to-hundred pc.

Most of the nuclear disk perturbations thus far identified in NUGA targets at scales of $\sim$a few
100\,pc are related to self-gravitating gas instabilities (e.g., in NGC\,4826; see
Fig.~\ref{Fig:1}). This finding is at odds with initial claims based on HST surveys of AGN spiral
hosts which suggested non self-gravitating perturbations are most likely linked with AGN feeding at
these scales (Martini \& Pogge \cite{mar99}). However, the new subarcsec NUGA maps have unveiled 
in some galaxies (NGC\,7217 (Combes et al \cite{paper2}), NGC\,6951 (Schinnerer et al 2004, in
prep.) and NGC\,4579 (Garc\'{\i}a-Burillo et al 2004, in prep.)) the presence of isolated GMC-like
gas units (of high Q) closely linked to the AGN source; these low mass gas concentrations,
suspicious of being directly responsible of AGN feeding, seem to coexist with the larger-scale
self-gravitating gas instabilities. Most remarkably, gas instabilities like the ones analyzed in the
two LINERs NGC\,4826 and NGC\,7217 may have, for quite different reasons, an {\it inhibiting} role
in AGN feeding.  Whether this puzzling scenario could be extrapolated to other low-L
AGN is being explored for the whole NUGA sample using diagnostic tools similar to the
ones developed in Papers I-II.

%
%

%
%
%

%
%

%
%


\end{document}